\documentclass{ws-jai}
\RequirePackage{doi}
\usepackage[flushleft]{threeparttable}

\usepackage{natbib}
\usepackage{hyperref}
\hypersetup{colorlinks=true,citecolor=blue,linkcolor=blue,urlcolor=blue,breaklinks=true}

\begin{document}

\catchline{}{}{}{}{} %

\markboth{Jean-Luc Margot}{A Data-Taking System for Planetary Radar Applications}

\title{A Data-Taking System for Planetary Radar Applications}

\author{Jean-Luc Margot$^{1,2}$}

\address{
$^{1}$Department of Earth, Planetary, \& Space Sciences, UCLA, Los Angeles, CA 90095, USA, jlm@epss.ucla.edu\\
$^{2}$Department of Physics \& Astronomy, UCLA, Los Angeles, CA 90095, USA, jlm@astro.ucla.edu
}

\maketitle

\corres{$^{1}$Corresponding author.}

\begin{history}
\received{October 23, 2020};
\revised{November 17, 2020};
\accepted{November 24, 2020};
\end{history}

\begin{abstract}
Most planetary radar applications require recording of complex
voltages at sampling rates of up to 20 MHz.  I describe the design and
implementation of a sampling system that has been installed at the
Arecibo Observatory, Goldstone Solar System Radar, and Green Bank
Telescope.  After many years of operation, these data-taking systems
have enabled the acquisition of hundreds of data sets, many of which
still await publication.
\end{abstract}


\section{Motivation}
\noindent Planetary radar astronomy is a discipline that enabled major
advances in our understanding of the scale of the Solar System, spin
states and surface properties of planets and satellites, and the
physical and dynamical properties of asteroids
\citep[e.g.,][]{ostr00,gior02,marg02,ches03,camp06,ostr06,marg07,tayl07}.
For reviews of the field, see \citet{ostr93,camp02,ostr02,benn15AIV}.
For an historical account of the development of the field up to 1996,
see \citet{butr96}.

The field is distinctive in part because essentially all results come
from only two transmitting facilities, the Arecibo Planetary Radar and
the Goldstone Solar System Radar.  The facilities are complementary in
that the former is $\sim$15 times more sensitive and the latter has
access to a fraction of the sky that is $\sim$2.5 larger
\citep{naid16}.  The facilities also probe slightly different depths
below the surface because Arecibo transmits primarily at S band (2380
MHz) whereas Goldstone transmits primarily at X band (8560~MHz).
Bistatic observations with reception at the 100 m Green Bank Telescope
(GBT) or other antennas also occur for a small fraction of the
observations.

Most planetary radar observations yield either 1D power spectra or 2D
range-Doppler images~\citep{evan68}, although 3D topographic maps can
be obtained in some circumstances \citep{marg99jgr}.  The sampling
rate requirements for power spectra are relatively modest ($<$60 kHz)
because the Doppler broadening of Solar System bodies at S band and X
band do not exceed 30 kHz, with the exception of Saturn's rings
\citep{nich05}.  For 2D imaging, the transmitted waveforms are most
commonly encoded with pseudo-noise (PN) binary phase codes or
modulated with linear frequency (chirp) ramps \citep{marg01chirp}.  In
both cases, the achievable range resolution is inversely proportional
to the bandwidth of the transmitted waveform.

NASA funded a major upgrade of the Arecibo
Planetary Radar in the late
1990s, which included a 1~MW S-band transmitter that can support
25~MHz bandwidth modulation \citep{gold96}.  %
In practice, 7.5~m resolution images are obtained by encoding the
transmitted waveform with a 0.05-$\mu$s-baud PN code, corresponding to
a bandwidth of 20 MHz.  However, the data acquisition systems
available after the upgrade were limited to 
sampling rates of 5 MHz, a factor of four lower than required for maximum
resolution.  This situation prompted the development of a baseband
recording system for use at Arecibo, GBT, and other receiving sites.
This data-taking system was subsequently deployed at Goldstone as well.

\section{Design Considerations}
Radar echoes in two orthogonal polarizations are detected by low-noise
receivers.  The receiver signals are amplified, filtered, and
typically converted to baseband prior to digitization for computer
storage and processing.  Important design considerations include the
resolution of the analog-to-digital (A/D) converter and the number of
bits that are retained for subsequent processing.  The data rate is
given by
\begin{equation}
d = 2 \times n_{\rm pol} \times n_{\rm bits} \times f_s,
\end{equation}
where $n_{\rm pol}$ is the number of recorded polarizations, $n_{\rm
  bits}$ is the number of recorded bits, $f_s$ is the sampling
frequency in Hz, and the factor of 2 indicates that both the in-phase
and quadrature components of the voltage signal are sampled, i.e., a
complex quantity.  For 8-bit sampling of 2 polarizations at 20 MHz,
the data rate is 640 Mb/s or 80 MB/s, which was prohibitive after the
Arecibo upgrade.

Fortunately, the first processing step in planetary radar imaging is a
range compression operation implemented as a digital correlation.
\citet{hage73} demonstrated that
digital correlations can be carried out effectively with two-, three-,
and four-level sampling of the input signals, resulting in
surprisingly low degradation of the signal-to-noise ratio (SNR)
compared to that obtained with a finely quantized signal.  For
instance, the ratio of the power obtained with a two-bit (four-level)
sampler to the power obtained with an ideal correlator exceeds 88\%
for a wide range of quantizer threshold settings \citep{koga98}.  In
particular, setting the sampling thresholds at zero and $\pm$0.996
times the standard deviation of the input voltage yields a correlator
efficiency of 88.12\% with integer output levels of \{-3, -1, +1, +3\}
\citep{schw86}.  The design requirement was therefore established at
2-bit sampling of 2 polarizations at 20 MHz, or a data rate of 20 MB/s.

The requirements for continuous sampling at low bit resolution are
unlike those of most scientific or industrial applications. Commercial
data acquisition products typically prize sampling resolutions of at
least 8 bits and provide no straightforward mechanism for retaining
only the most significant bits.  This realization led to a
custom-built design and implementation for what became known as the
Portable Fast Sampler (PFS), with available sampling modes listed in
Table~\ref{tab-modes}.

\begin{table}[h]
  \begin{center}
\begin{tabular}[h]{|r|r|r|r|r|}
\hline
mode & channels & bits & sampling rate & data rate \\
\hline
         0 &  2 & 1 & (N/A) & \\
         1 &  2 & 2 & 5 -- 40 MHz  &    2.5 -- 20 MB/s\\
         2 &  2 & 4 & 5 -- 20 MHz  &    5   -- 20 MB/s\\ 
         3 &  2 & 8 & 5 -- 10 MHz  &    10  -- 20 MB/s\\
         4 &  4 & 1 & (N/A) & \\
         5 &  4 & 2 & 5 -- 20 MHz  &    5   -- 20 MB/s\\
         6 &  4 & 4 & 5 -- 10 MHz  &    10  -- 20 MB/s\\
         7 &  4 & 8 & (N/A) & \\
\hline
\end{tabular}
\end{center}
\caption{Description of sampling modes.}
\label{tab-modes}
\end{table}

\section{Hardware Implementation}

The sampling requirement is achieved with two Analog Devices
AD9059/PCB, which are dual-channel 8-bit A/D boards with maximum
conversion rates of 60 million samples per second.  These boards
require only a +5~V power supply and a TTL-compatible encode clock.
Full-scale on the A/D converters is achieved for input signals of 1 V
peak-to-peak driving a 50 $\Omega$ termination. Therefore, input
voltages with standard deviations near 0.25~V (1 dBm) yield optimum
two-bit sampling.  Digital outputs are TTL compatible.

In order to emulate 2-, 4- or 8-bit sampling, the 32 A/D digital
outputs are connected to a programmable logic device (PLD) for bit
selection and packing.  A device from the Altera MAX 7000S family
(EPM7128SLC84-10) with plastic J-lead chip carrier (PLCC) 84-pin
packaging is used for this purpose.  It requires a +5~V power supply
and accepts TTL-level input voltages.  The supply voltages for
internal logic and input buffers (VCCINT) and for output drivers
(VCCIO) are both set to 5~V.  The selected bits are packed into
sixteen data bits that are sent to four differential line drivers
(SN75ALS194) for RS-422 transmission over a 7-foot-long twisted pair
cable (EDT CAB-AA 016-00427-00).

A parallel interface card (EDT PCI CD-20) is used to transfer the data
to computer memory and storage disks at rates of up to 20 MB/s.  The
interface card is connected to the PCI bus of a rack-mounted computer
that runs the Linux operating system.  The interface card can handle
four general-purpose control outputs (FUNCT0--3), which are used to
specify the sampling mode
(written to FUNCT1--3) and to toggle a data acquisition enable signal
(written to FUNCT0).

Two station clock signals phase-locked to a high-accuracy frequency
standard are required.
The first signal is a one pulse-per-second (PPS) TTL signal, which is
used to trigger the start of the data acquisition.  The second signal
is a +13 dBm ($\sim$3 Vpp into 50 Ohms) sampling clock of arbitrary
frequency, although most observations are conducted with a 5, 10, or
20 MHz sampling clock.  The sampling clock is a simple sine wave that
is converted to a TTL square wave and distributed to the A/D boards
and PLD.

The enclosure for the A/D boards, PLD, and associated electronics is a
Lansing Instrument Corp. B2F18-001C, which is 2 rack units (3.47~in),
full rack width (16.73~in), and 18~in deep.  BNC connectors are used
to connect the 1 PPS, sampling clock, and four analog input signals.

\section{Software Implementation}

A computer program ({\tt pfs\_radar}) written in C can be executed
from the command line or graphical user interface to control the
operation of the data-taking system.  Command-line options enable
specification of the sampling mode ({\tt -m}), recording duration
in seconds ({\tt -secs}), and start time ({\tt -start
  yyyy,mm,dd,hh,mm,ss}) specified in Universal Time.  The computer
clock is synchronized to accurate time servers with the Network Time
Protocol (NTP).

Planetary radar observations require extreme ($<$10 ns) timing
precision.  Upon execution of the data-taking program, the computer
allocates ring buffer memory, opens log and data files, then suspends
execution until 0.5~s before the desired start time.  At that point,
the data acquisition enable bit (FUNCT0) is activated, which signals
to the PLD that data acquisition will commence on the upcoming rising
edge of the 1 PPS signal.

When the rising edge of the 1 PPS is detected while the data
acquisition bit is enabled, the PLD toggles the Input Data Valid (IDV)
bit, which instructs the 16-bit interface card to transfer data to PCI
bus memory.  The PLD is programmed to generate a Receive Timing (RXT)
signal for use by the interface card, which stores inputs only at the
rising edge of the RXT signal.  Data are transferred 16 bits at a time
until the C program detects that the end time of data-acquisition has
been reached.  At that point, the data acquisition bit is deactivated,
ring buffers are cleared, the log file is flushed, and the whole
process can start over.

Data files contain exclusively the raw data in packed format.  Each
data file name encodes the receiver start time in the format {\tt
  datayyyymmddhhmmss}.  Ancillary information about sampling mode,
buffer sizes, data rates, etc. are stored in the separate ASCII log
file.  Data-taking progress can be monitored remotely by examining
this log file.

Most radar observations proceed with multiple transmit/receive cycles,
where the timing of the receive cycles can be calculated in advance.
Many observations use a simple script that repeatedly invokes the
data-taking program with the value of the next receive start time.

Ancillary programs enable data inspection and analysis in near real
time.  One can display histograms and statistics of the input data
with {\tt pfs\_hist} and {\tt pfs\_stats}.  Unpacking and downsampling
of the data can be done with the programs {\tt pfs\_unpack} and {\tt
  pfs\_downsample}.  Spectral analysis is facilitated with {\tt
  pfs\_fft}.  The data-taking software and all ancillary programs are
available on GitHub at
\href{https://github.com/UCLA-RADAR-Group/pfs}{https://github.com/UCLA-RADAR-Group/pfs}.

\section{Applications}

The data-acquisition system was installed at Arecibo (2 units), the
Green Bank Telescope (2 units), and Goldstone (4 units).  It has been
used to acquire most of the radar echoes at Arecibo (2000--present),
Goldstone (2001--2014), and Green Bank (2001--2017).

The system was used for the first radar detection of a Solar System
object at NASA's Deep Space tracking station (DSS-63) in Madrid,
Spain. The asteroid 6489 Golevka was detected on 4 June 1999 with the
expected Doppler bandwidth and at the expected frequency.
Observations of the same asteroid were also obtained at
Arecibo in 2003 and led to the first detection of the Yarkovsky
orbital drift~\citep{ches03}.

\begin{figure}[h]
	\centering
	\includegraphics[angle=0,width=6.5in]{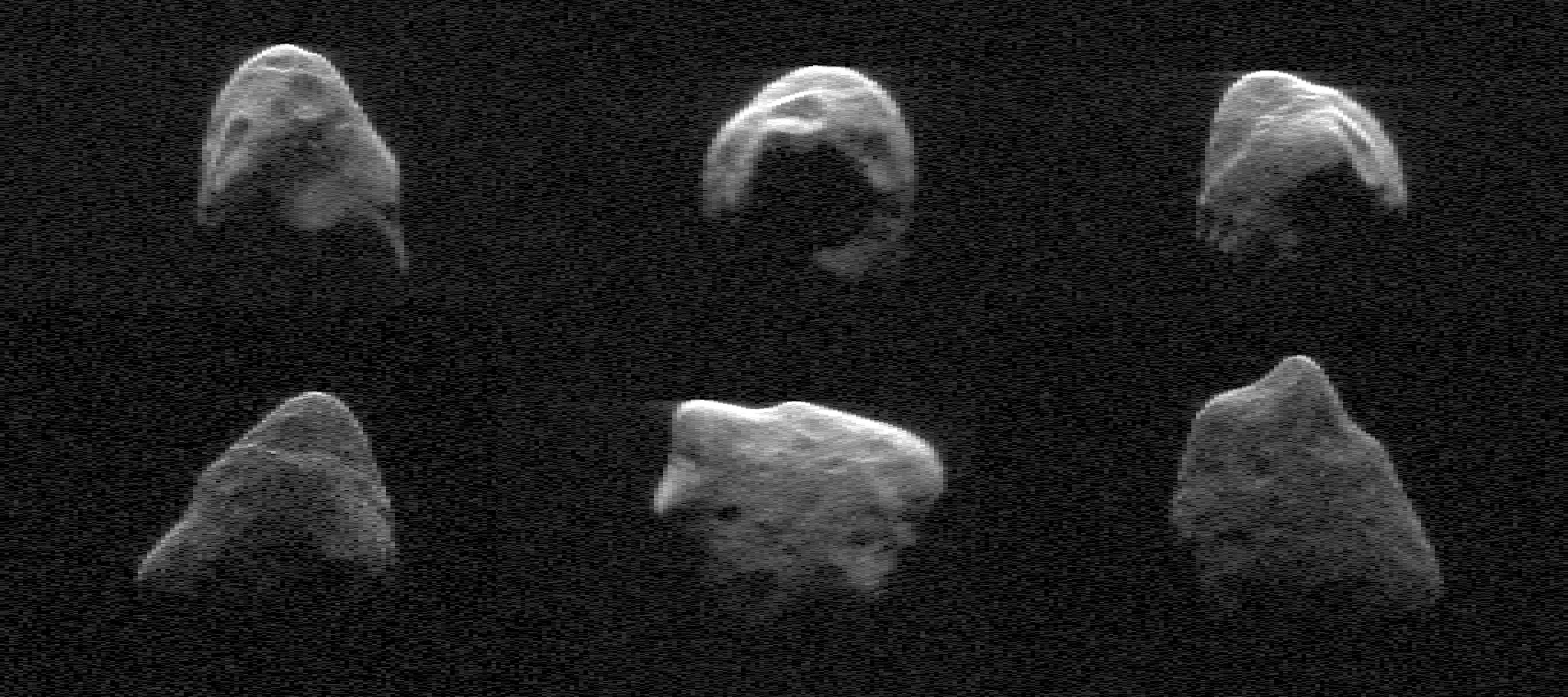}
	\caption{This sequence of 15~m resolution Arecibo radar images
          show asteroid 1999 JM8 on consecutive days.  }
\label{fig-jm8} 
\end{figure}

Arecibo radar images of asteroid 1999 JM8 were obtained at 15~m
resolution with the data-taking system on 1--9 August 1999.  At the
time, these images were the highest resolution images of an asteroid
ever obtained (Figure~\ref{fig-jm8}).  This distinction was lost when
the NEAR-Shoemaker spacecraft lowered its orbit sufficiently close to
asteroid 433 Eros.  Analysis of the radar data revealed that
the asteroid
has an effective diameter of 7~km and a non-principal-axis
rotation with a dominant periodicity near 7 days \citep{benn02}.

\begin{figure}[h]
	\centering
	\includegraphics[angle=0,width=6.5in]{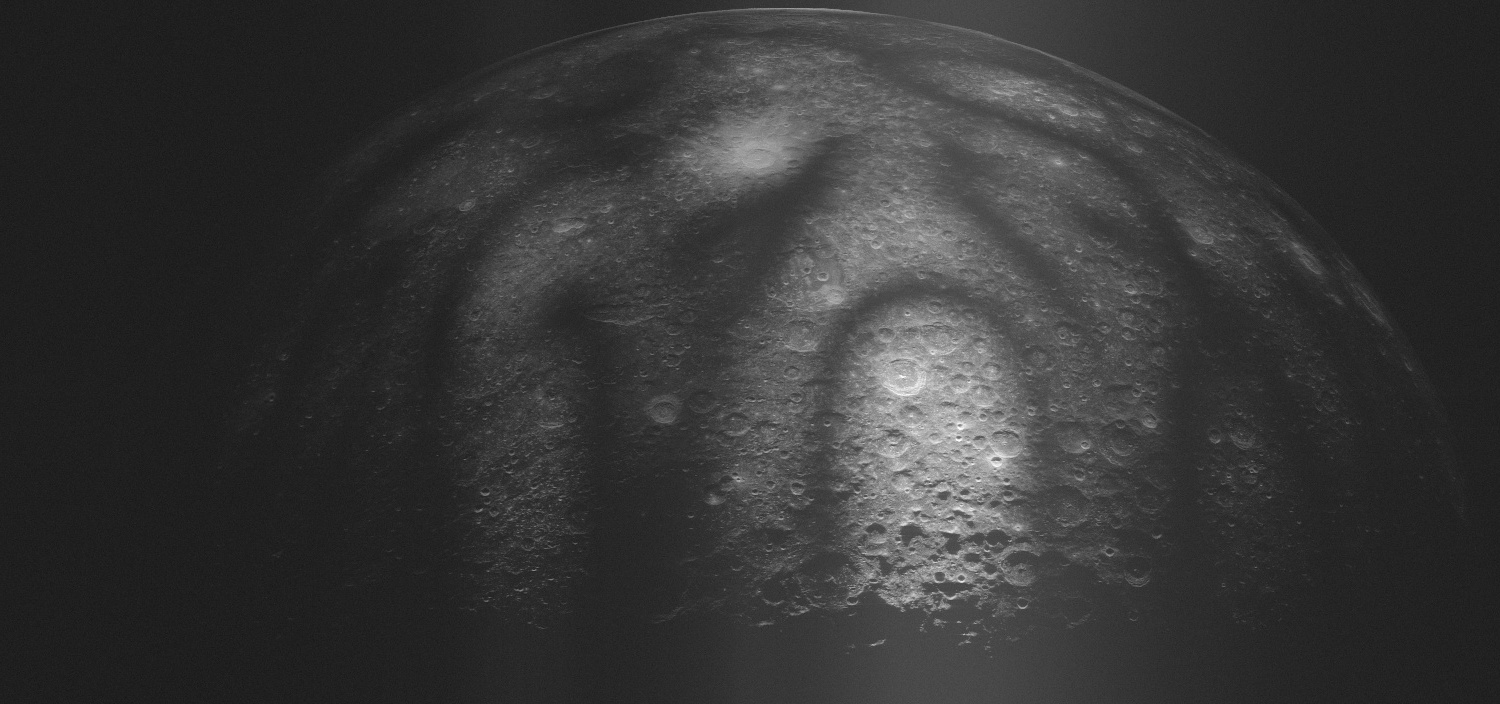}
	\caption{This Arecibo to St-Croix radar image shows the beam
          pattern of the Arecibo telescope painted on the surface of
          the Moon. The main lobe and first null are visible
          around the south polar region. Echoes from the rest of the
          Moon appear through the sidelobes of the beam pattern.  }
\label{fig-moon} 
\end{figure}

Observations of the Moon with Arecibo transmitting at 2380 MHz and the
25 m VLBA antenna in St-Croix receiving were conducted on 19--21
November 2000.  The data-taking system was used to record images at
30~m resolution (Figure \ref{fig-moon}).

The data-taking system enabled
some of the first scientific observations at the Green Bank Telescope.
Arecibo-GBT radar images of asteroid 2001 EC16 and Venus
at 15~m and 150--300 m resolutions, respectively, were obtained on
24--26 March 2001.
Because the round-trip light-time to the asteroid was 11~s, and
because it takes several seconds to switch between transmit and
receive modes, monostatic observations would have limited the
frequency resolution to the reciprocal of the $\sim$5~s receive time,
approximately 0.2~Hz.  Continuous transmission from Arecibo and
continuous reception at the GBT for several minutes enabled the
acquisition of spectra with two orders of magnitude better frequency
resolution.  One goal of the Venus observations was to obtain
topographic maps with a 3D interferometric technique~\citep{marg99a}.

Other notable radar observations from Arecibo include Saturn's rings
in 2001, 2001, and 2003 \citep{nich05}, binary asteroid 1999 KW4 at
7.5~m resolution in 2001 \citep{ostr06}, and many observations of the
Moon~\citep[e.g.,][]{camp07}.

The sampling system has been used in a Goldstone-GBT configuration to
produce high-precision measurements of planetary spin states.  The
observations revealed that Mercury has a liquid outer core
\citep{marg07} and enabled a measurement of the size of its core
\citep{marg12jgr}.  They also enabled the first measurement of the
spin precession rate of Venus and revealed that Venus exhibits
length-of-day variations of tens of minutes \citep{marg21}.

\section*{Acknowledgments}
JLM is funded in part by NASA grants 80NSSC18K0850 and 80NSSC19K0870.
I thank Jeff Hagen and Joseph Jao for software contributions.  The
data-taking system was initially funded by the National Astronomy and
Ionosphere Center.



\end{document}